\documentclass[conference]{IEEEtran}
\IEEEoverridecommandlockouts

\def\BibTeX{{\rm B\kern-.05em{\sc i\kern-.025em b}\kern-.08em
    T\kern-.1667em\lower.7ex\hbox{E}\kern-.125emX}}

\usepackage{cite}
\usepackage{times}

\usepackage{xcolor}
\usepackage[hidelinks]{hyperref}
\usepackage{url}
\usepackage[T1]{fontenc}
\usepackage{diagbox}
\usepackage{amsmath}
\usepackage{amssymb}
\usepackage{mathtools}
\usepackage{enumitem}
\usepackage{color}
\usepackage[caption=false, font=footnotesize]{subfig}
\usepackage{multirow}

\usepackage{pifont}

\usepackage{makecell}
\usepackage{todonotes}
\usepackage[bottom]{footmisc}

\usepackage{datetime}
\usepackage[utf8]{inputenc}
\usepackage{csquotes}
\usepackage{listings}
\usepackage[skip=0pt,font=small]{caption}
\usepackage{comment}
\usepackage{enumitem}

\usepackage{subfig}
\usepackage{graphicx}
\usepackage{xcolor}
\usepackage{colortbl}
\usepackage{lettrine}

\usepackage{algorithm}
\usepackage{algorithmicx}  
\usepackage{algpseudocode} 
\usepackage{listings}

\usepackage{textcomp}
\usepackage{tikz}
\usepackage{soul}
\usepackage{adjustbox}
\usepackage{pifont}

\usepackage{mwe}
\usepackage{booktabs}
\usepackage{ulem}

\usepackage[switch]{lineno}

\newcolumntype{?}{!{\vrule width 1pt}}
\newcolumntype{^}{!{\vrule width 1.2pt}}

\newcommand{\falcongemm}{{\textsc{FalconGEMM}}}

\definecolor{myblue}{RGB}{91,155,213}
\definecolor{mydark}{RGB}{0,0,0}
\newcommand*\circled[1]{\kern-2.5em%
  \put(0,4){\color{black}\circle*{18}}\put(0,4){\circle{16}}%
  \put(-3,0){\color{white}\bfseries\large#1}~~}

\newcommand{\appropto}{\mathrel{\vcenter{
  \offinterlineskip\halign{\hfil$##$\cr
    \propto\cr\noalign{\kern2pt}\sim\cr\noalign{\kern-2pt}}}}}

\usepackage{enumitem}
\usepackage{adjustbox}
\DeclareRobustCommand*\circled[1]{\tikz[baseline=(char.base)]{
            \node[shape=circle,fill,inner sep=1.3pt] (char) {\textcolor{white}{#1}};}}

\newcommand{\squishlist}{
\begin{list}{$\bullet$}
 { \setlength{\itemsep}{0pt}
   \setlength{\parsep}{3pt}
   \setlength{\topsep}{3pt}
   \setlength{\partopsep}{0pt}
   \setlength{\leftmargin}{1.2em}
   \setlength{\labelwidth}{1em}
   \setlength{\labelsep}{0.6em}
 }
}

\newcommand{\squishend}{
 \end{list}
}

\begin{document}

\title{FalconGEMM: Surpassing Hardware Peaks with  Lower-Complexity Matrix Multiplication}
\makeatletter
\DeclareRobustCommand*{\IEEEauthorrefmark}[1]{%
  \raisebox{0pt}[0pt][0pt]{\textsuperscript{\footnotesize\ensuremath{\@arabic#1}}}%
}
\makeatother
\author{
    \IEEEauthorblockN{
        Honglin Zhu\IEEEauthorrefmark{1},
        Jiaping Cao\IEEEauthorrefmark{2},
        Jiang Shao\IEEEauthorrefmark{3},
        Siyuan Feng\IEEEauthorrefmark{4},
        Qian Qiu\IEEEauthorrefmark{5},
        Peng Chen\IEEEauthorrefmark{6}, \\
        Xu Zhang\IEEEauthorrefmark{7},
        Yixian Zhou\IEEEauthorrefmark{8},
        Man Lung Yiu\IEEEauthorrefmark{9},
        Guang Ji\IEEEauthorrefmark{10},
        Minwen Deng\IEEEauthorrefmark{11},
        Jintao Meng\IEEEauthorrefmark{12},
        Wenxi Zhu\IEEEauthorrefmark{13},
    }

    \IEEEauthorblockA{
        \IEEEauthorrefmark{1,2,5,8,11,13}Tencent, Shenzhen, China  \quad
        \IEEEauthorrefmark{2,9}The Hong Kong Polytechnic University, Hong Kong   \\
        \IEEEauthorrefmark{3,10}NVIDIA, Beijing, China \quad
        \IEEEauthorrefmark{4}Shanghai Innovation Institute, 
        Shanghai, China  \quad
        \IEEEauthorrefmark{6}RIKEN, Tokyo, Japan \\
        \IEEEauthorrefmark{7}Southern University of Science and Technology, Shenzhen, China \\
        \IEEEauthorrefmark{7,12}Shenzhen Institute of Advanced Technology, Chinese Academy of Sciences, Shenzhen, China 
    }

    \IEEEauthorblockA{
        \IEEEauthorrefmark{1}honglinzhu@tencent.com,
        \IEEEauthorrefmark{2}vincecao@tencent.com,
        \IEEEauthorrefmark{3}jiangs@nvidia.com,
        \IEEEauthorrefmark{4}syfeng@sii.edu.cn, 
        \IEEEauthorrefmark{5}qianqiu@tencent.com, \\ 
        \IEEEauthorrefmark{6}peng.chen@a.riken.jp, 
        \IEEEauthorrefmark{7}xu.zhang3@siat.ac.cn, 
        \IEEEauthorrefmark{8}walnuts0715@gmail.com, 
        \IEEEauthorrefmark{9}csmlyiu@comp.polyu.edu.hk, \\ 
        \IEEEauthorrefmark{10}gji@nvidia.com,
        \IEEEauthorrefmark{11}danierdeng@tencent.com,
        \IEEEauthorrefmark{12}jt.meng@siat.ac.cn,
        \IEEEauthorrefmark{13}wenxizhu@tencent.com
    }

\thanks{
    Honglin Zhu and Jiaping Cao contributed equally to this work. 
    Wenxi Zhu and Jintao Meng are the corresponding authors.
}

}

\maketitle

\thispagestyle{plain}
\pagestyle{plain}

\begin{abstract}
\color{black}
Peak breaking Matrix Multiplication is a promising technique to improve the performance of DL, especially in LLM training and inference. We present FalconGEMM, a cross-platform framework that automates the deployment, optimization, and selection of Lower-Complexity Matrix Multiplication Algorithms (LCMAs) across diverse hardware. There are three key innovations: (1) a Deployment Module that enables portable execution across various hardware and input configurations through code generation; (2) an Execution Module with Group-Parallel Optimizations that maximizes on-chip data reuse, utilizes parallel resources, and reduces bandwidth overhead; and (3) a Decision Module featuring a lightweight analytical performance model to select the optimal strategy based on matrix shapes and hardware profiles. Extensive evaluation is conducted on LLM workloads across GPU (H20, A100) and CPU (ARM, x86) architectures with multiple data types. FalconGEMM succeeds in delivering peak breaking performance and outperforms GEMM libraries (e.g., cuBLAS, CUTLASS, Intel MKL, etc) by 7.59\%–17.85\% and LCMA competitors like AlphaTensor by 12.41\%–55.61\%. Our framework makes the theoretical promise of LCMAs practical for production deployment across the heterogeneous landscape of modern hardware.
\end{abstract}

\section{Introduction}
\label{sec:intro}
General Matrix Multiplication (GEMM) serves as the computational backbone for modern deep learning (DL) and high-performance computing (HPC) workloads, dominating the runtime of architectures ranging from Transformers to Convolutional Neural Networks (CNNs).
As model sizes scale exponentially, the $O(N^3)$ complexity of standard GEMM algorithms has become a critical bottleneck.
To address this, a class of fast matrix multiplication algorithms has been developed. Collectively, we term them Lower-Complexity Matrix Multiplication Algorithms (LCMAs), such as Strassen's algorithm~\cite{strassen1969gaussian}, Laderman's algorithm~\cite{laderman1990new}, and algorithms discovered by AlphaTensor~\cite{fawzi2022discovering}.
By decomposing matrices into submatrices and performing linear combinations among them, LCMAs reduce the asymptotic complexity to $O(N^{\log_2 7})$ or lower.
Since vendor-optimized libraries like cuBLAS\cite{nvidia2025cublas}, Intel MKL\cite{intel2025mkl}, and localized optimizations~\cite{deepseek2025deepgemm, nvidia2025cutlass} have already pushed standard GEMM performance close to hardware peaks, further acceleration must rely on algorithmic innovations like LCMAs.
Theoretically, this reduction offers the potential to accelerate large-scale model training and inference beyond the physical limits of current hardware. 
However, due to the extra memory overhead and lack of platform portability, few implementations can work efficiently as a backend to improve the performance of DL.

Translating the theoretical arithmetic reduction of LCMAs into practical speedups presents three major challenges:
\circled{1} ~\textit{Lack of Cross-Platform Portability}: LCMAs introduce complex data dependencies and linear combination steps that differ significantly from standard GEMM. Implementing LCMAs requires writing highly specialized programs with complex data dependencies and also handling potential race conditions, which is likely to be inefficient.
Besides, manually optimizing for every hardware backend (e.g., NVIDIA Tensor Cores, Arm NEON, x86 AVX) has a vast implementation space, leading to a lack of portable solutions.
\circled{2}~\textit{Memory Overhead}: The multi-phase nature of LCMAs necessitates the materialization of large intermediate tensors. Writing these intermediate results to off-chip memory and reading them back for final accumulation consumes substantial memory bandwidth. This additional memory traffic often saturates the bus, negating the computational gains derived from the algorithm's lower arithmetic complexity.
\circled{3}~\textit{Difficulty in Decision}: LCMAs are not universally faster; their benefit depends heavily on matrix shape, hardware bandwidth, and compute throughput. The performance cutoff point where the reduction in arithmetic operations outweighs the overhead is difficult to predict. It is important for users to have a lightweight mechanism to determine whether an LCMA will provide a speedup and which specific algorithm is optimal for their specific workload.

\falcongemm{} is developed to effectively bridge the gap between the theoretical promise of LCMAs and the practical demands of production-level Deep Learning. By systematically addressing the complexities of portability, memory overhead, and algorithm selection, the framework transforms "peak-breaking" performance from a mathematical possibility into a cross-platform reality.
The core contribution of \falcongemm{} lies in its three-pillar architecture:
{
    \vspace{-3pt}
    \setlength{\leftmargini}{10 pt}
    \begin{itemize}%[label=(\Roman*)]

 \item \textbf{Deployment:} We abstract LCMAs and introduce a four-stage fast implementation; then, we apply code generation to automatically address the \textit{complexity of portability} with millions of implementation variants.

 \item \textbf{Execution:}  We propose a Group-Parallel Optimization to eliminate \textit{memory overhead} with a fused workflow, and address the induced parallel granularity and cache thrashing with Split-Group Parallelism and Cache-Aware Scheduling.

\item  \textbf{Decision:} We conduct a theoretical arithmetic intensity analysis using a lightweight performance model, and apply it to guide optimal \textit{algorithm selection} between multiple LCMAs and GEMM.

\end{itemize}
}
Extensive evaluation is conducted on five computing devices, including both GPU (H20, A100) and CPU (ARM, x86) with multiple data types. \falcongemm{} outperforms state-of-the-art GEMM libraries by 7.59\%--17.85\% (mostly beyond the peak) and LCMA implementations by 12.41\%--55.61\%. 
When using \falcongemm{} as PyTorch's backend, we achieve 11.46\%-18.12\% average performance gains in the prefill stage when inferencing HunyuanVideo, Qwen3.5 and  DeepSeek-R1. {\color{black}The step-wise evaluation confirms a stable optimization with at most 7.18\% extra speedup by the Execution Module. The roofline analysis shows that \falcongemm{} succeeds in delivering optimal and stable peak breaking performance with the Decision Module, with about 19.31\% and 11.37\% performance gains compared to standard GEMM and Strassen's algorithm, respectively.

\section{Definition and Motivation}\label{sec:background}
We first define Lower-Complexity Matrix Multiplication Algorithms (LCMAs), then discuss our motivation and intuition.

\begin{table}[b]
\scriptsize
\centering
\caption{Notations of LCMA}
\label{tbl:lcma_notation}
\setlength{\tabcolsep}{10pt}
\begin{tabular}{|c|l|}\hline
Notation & Meaning \\ \hline \hline
$A \in \mathbb{R}^{M \times K}, B \in \mathbb{R}^{K \times N}$ & Input matrices \\ \hline
$C \in \mathbb{R}^{M \times N}$ & Output matrix, $C=AB$ \\ \hline
$\mathcal{L}=\langle m,k,n, R, U,V,W\rangle$ & Definition of an LCMA \\ \hline
$m, k, n$ & Grid dimensions for partitioning $M, K, N$ \\ \hline
$R$ & Rank (number of multiplications) \\ \hline
$A_{i,\ell}$ & Submatrix of $A$, for $i \in [1, m], \ell \in [1, k]$ \\ \hline
$B_{\ell,j}$ & Submatrix of $B$, for $\ell \in [1, k], j \in [1, n]$ \\ \hline
$C_{i,j}$ & Submatrix of $C$, for $i \in [1, m], j \in [1, n]$ \\ \hline
$\tilde{A}_r$ & Matrix combined by $A$, for $r \in [1, R]$ \\ \hline
$\tilde{B}_r$ & Matrix combined by $B$, for $r \in [1, R]$ \\ \hline
$H_r$ & Intermediate matrix, for $r \in [1, R]$ \\ \hline
$U, V, W$ & Coefficient tensors in $\{-1,0,1\}$ \\ \hline
\end{tabular}
\end{table}

\subsection{Definition of LCMA}
\label{sec:lcma_definition}

Given two matrices $A \in \mathbb{R}^{M \times K}$ and $B \in \mathbb{R}^{K \times N}$ ($\mathbb{R}$ can be represented with Int8 and floating point types such as FP8, FP16, BF16, or FP32),  
\textbf{Lower-Complexity Matrix Multiplication Algorithms} (LCMAs) are a class of algorithms that can be applied to compute matrix multiplication Equation (\ref{eq:lcma}) by performing fewer multiplications than the standard algorithm: 
\begin{equation}
\label{eq:lcma}
A \times B = C, \quad \quad C \in \mathbb{R}^{M \times N}.
\end{equation}
Specifically, an LCMA is defined by a tuple $\mathcal{L} = \langle m, k, n, R, U, V, W \rangle$, where $\langle m, k, n \rangle$ indicates the algorithm grid dimensions, $R$ is the rank of the algorithm, and the algorithm coefficient tensors are %$U \in \mathbb{R}^{R \times m \times k}$, $V \in \mathbb{R}^{R \times k \times n}$ and $W \in \mathbb{R}^{R \times m \times n}$ (the entry values of coefficients are typically restricted to a small discrete set (e.g., $\{-1, 0, 1\}$) to avoid floating-point arithmetic issues).
$U \in \mathbb{S}^{R \times m \times k}$, $V \in \mathbb{S}^{R \times k \times n}$  and $W \in \mathbb{S}^{R \times m \times n}$, here $\mathbb{S} = \mathbb{R}$. In most cases $\mathbb{S} = \{-1, 0, 1\}$~\cite{schwartz2023pebbling, moran2026complex}.
LCMA first partitions $A$, $B$ and $C$ into grids of the following submatrices corresponding to the grid dimensions of $m, k, n$, then the Eq.~\ref{eq:lcma} can be written as:
\begingroup
\setlength{\arraycolsep}{3pt} 
\begin{equation}
\label{eq:lcma_grid}
\tiny
\begin{aligned}
\begin{pmatrix}
A_{1,1} & \cdots & A_{1,k} \\
\vdots & A_{i,\ell} & \vdots \\
A_{m,1} & \cdots & A_{m,k}
\end{pmatrix}
\times
\begin{pmatrix}
B_{1,1} & \cdots & B_{1,n} \\
\vdots & B_{\ell,j} & \vdots \\
B_{k,1} & \cdots & B_{k,n}
\end{pmatrix}
=
\begin{pmatrix}
C_{1,1} & \cdots & C_{1,n} \\
\vdots & C_{i,j} & \vdots \\
C_{m,1} & \cdots & C_{m,n}
\end{pmatrix}
\end{aligned}
\end{equation}
\endgroup
here $A_{i,l} \in \mathbb{R}^{\lceil \frac{M}{m} \rceil \times \lceil \frac{K}{k} \rceil}$, $B_{l,j} \in \mathbb{R}^{\lceil \frac{K}{k} \rceil \times \lceil \frac{N}{n} \rceil}$, and $C_{i,j} \in \mathbb{R}^{\lceil \frac{M}{m} \rceil \times \lceil \frac{N}{n} \rceil}$.

The standard multiplication algorithm performs $m \cdot k \cdot n$ multiplications between submatrices, whereas LCMA uses only $R$ matrix multiplications (where $R < m \cdot k \cdot n$). The above computation for Eq.~\ref{eq:lcma_grid} will be partitioned into four stages. 
\begin{enumerate}
 \item \textbf{Combine A:} 
\begin{equation}
\label{eq:addA}
    \tilde{A}_r =  \sum_{i=1}^{m} \sum_{\ell=1}^{k} U_{r,i,\ell} \cdot A_{i,\ell}, \quad \text{where } %U_{r,i,\ell} \in \mathbb{S},  
    1 \leq r \leq R 
\end{equation}
 
 \item \textbf{Combine B:}   
\begin{equation}
\label{eq:AddB}
    \tilde{B}_r = \sum_{\ell=1}^{k} \sum_{j=1}^{n} V_{r,\ell,j} \cdot B_{\ell,j}, \quad \text{where } %V_{r,\ell,j} \in \mathbb{S}, 
    1 \leq r \leq R  
\end{equation}
 
 \item \textbf{GEMM:} 
\begin{equation}
\label{eq:BatchedGEMM}
    H_r = \tilde{A}_r \cdot \tilde{B}_r, \quad \text{where } 1 \leq r \leq R 
\end{equation}
 
\item \textbf{Combine H:}  
\begin{equation}
\small
\label{eq:addH}
\begin{split}
    C_{i,j} = \sum_{r=1}^{R}  W_{r,i,j} \cdot H_r, 
    \text{where } 
    1\leq i\leq m, 1\leq j\leq n 
\end{split}
\end{equation}
\end{enumerate}
Here in \textbf{Combine A} and \textbf{Combine B} stage, Eq.~\ref{eq:addA} and \ref{eq:AddB} utilizes the coefficient tensors $U$ and $V$ to linearly combine submatrices $A$ and $B$ into $\{\tilde{A}_r\}_{r=1}^R$ and $\{\tilde{B}_r\}_{r=1}^R$, respectively. 
Then in \textbf{GEMM} stage,  there are $R$ multiplications between  $\tilde{A}_r$ and $\tilde{B}_r$  to generate $R$ intermediate matrices $H_r$ with Eq.~\ref{eq:BatchedGEMM}.
Finally \textbf{Combine H} uses coefficient tensor $W$ to linearly combine intermediate $\{H_r\}_{r=1}^R$ to obtain the final result submatrices $\{C_{i,j}\}_{i=1j=1}^{m\ \ n}$ according to Eq.~\ref{eq:addH}. All the symbols used in this paper are  summarized in Table~\ref{tbl:lcma_notation}.

\begin{figure}[tbp]
    \centering
    \includegraphics[width=1\linewidth]{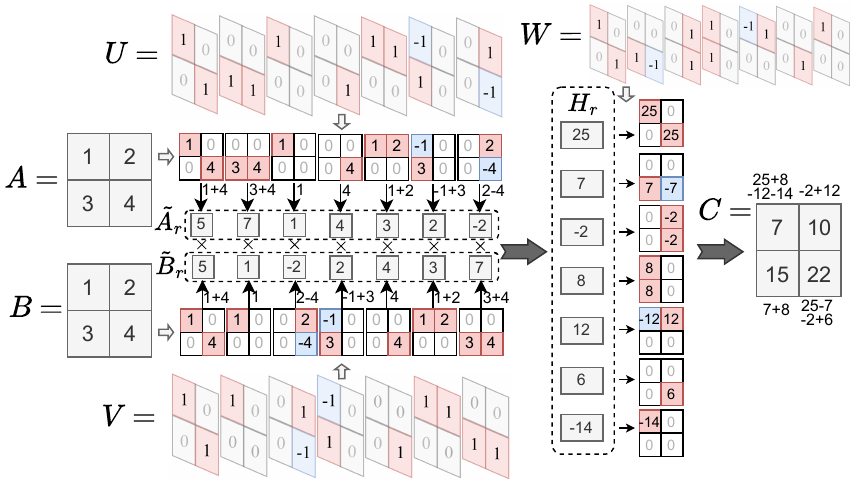}
    \caption{Dataflow of Strassen's algorithm ($\langle 2,2,2\rangle, R=7$). The figure visualizes its algorithm coefficient tensors $U, V, W$ required to compute the linear combinations and the $R=7$ multiplications. This specific example illustrates the case where $M=N=K=2, m=n=k=2$, but LCMA can generalize to arbitrary sizes. }
    \label{fig:strassen_flow}
\end{figure}

As an example, Strassen's algorithm~\cite{strassen1969gaussian} is one of the most well-known LCMA with grid dimensions $m = n = k = 2$ and rank $R = 7$ (denoted as ($\langle 2,2,2 \rangle, R=7$)).  Figure~\ref{fig:strassen_flow} illustrates its dataflow for multiplication of matrices $A$ and $B$. For simplicity, we demonstrate a special case where each submatrix of $A$ and $B$ contains a single scalar. The coefficient tensors $U$ and $V$ define how the input submatrices are combined to form seven intermediate matrices ($H_1, \dots, H_7$). Subsequently, the tensor $W$ defines how these seven intermediate matrices are linearly combined to produce the four output submatrices.  

We should also note that this definition can apply to all LCMA configurations discovered to date and is adopted throughout this paper.  
For example, other classical algorithms include Laderman's algorithm~\cite{laderman1990new} ($\langle 3,3,3 \rangle, R=23$) and the two-level recursive Strassen's algorithm~\cite{strassen1969gaussian} ($\langle 4,4,4 \rangle, R=49$).  
More recently, AlphaTensor~\cite{fawzi2022discovering} has additionally discovered a large number of novel LCMAs, such as ($\langle 3,4,5 \rangle, R=47$) and ($\langle 4,4,5 \rangle, R=63$).

\subsection{Motivation}
\label{sec:related}

While these LCMAs theoretically reduce arithmetic complexity, achieving actual performance gains in practical deployment is nontrivial, and existing works still have limitations.

\paragraph{Platform Portability}
It is challenging to efficiently deploy various LCMAs on diverse hardware.  
We model this complexity with a search space $\mathcal{N}_{impl}$ as a combination of the following configurations:
\begin{equation}
\label{eq:impl_complexity}
    \mathcal{N}_{impl} \approx \mathcal{N}_{algo} \times \mathcal{N}_{hw} \times \mathcal{N}_{dtype} \times \mathcal{N}_{config}
\end{equation}
here $\mathcal{N}_{algo}$ is defined as the number of different LCMAs. The discoveries by AlphaTensor~\cite{fawzi2022discovering} have expanded the set of viable algorithms ($\mathcal{N}_{algo}$) to the order of $10^2$. 
$\mathcal{N}_{hw}$ and $\mathcal{N}_{dtype}$ are defined as  the available varieties of hardware and data types used by modern deep learning frameworks. They span multiple generations of NVIDIA, AMD, Intel, and ARM architectures ($\mathcal{N}_{hw} \approx 10$) and data types ($\mathcal{N}_{dtype} \approx 5$) among FP32, FP16, BF16, INT8, etc.
$\mathcal{N}_{config}$ is termed the tuning configurations 
involving tile sizes, warp shapes, and pipeline stages. We anticipate the number of tuning combinations to be $\mathcal{N}_{config} \approx 10^3$.  
The resulting product $\mathcal{N}_{impl}$ is in the millions of potential implementation variants. This implementation burden acts as a prohibitive barrier to widespread adoption. Previous LCMAs deployment~\cite{huang2017generating,benson2015framework} still struggles to provide a highly optimized framework across both diverse hardware and LCMAs.

\paragraph{Parallel Partitioning}
Existing implementations~\cite{benson2015framework, huang2016strassen, huang2020strassen} typically decompose LCMA into $R$ tasks, each computing one $H_r$ from submatrices of $A$ and $B$ and then combining the product into several $C_{i,j}$. This $H_r$-parallel approach suffers from three inefficiencies.  
\circled{1} {Redundant memory access} on repeatedly loading submatrices of $A$ and $B$ occurs during the computation of $A_r$ and $B_r$~\cite{benson2015framework, huang2016strassen, huang2020strassen} . 
\circled{2} {Operator fragmentation} happens when decomposing the GEMM stage into $R$ smaller GEMMs, which degrades resource utilization, occupancy, and latency hiding~\cite{dodovic2026fusion} compared to a single batched GEMM~\cite{huang2020strassen, wang2025kami, dodovic2026fusion, he2025stragcn}.  
\circled{3} {Write Conflicts} is introduced because each $C_{i,j}$ depends on multiple $H_r$  (e.g.,  $C_{1,1}$ in Strassen's Algorithm depends on $H_1$, $H_4$, $H_5$, and $H_7$). 
Separating conflicting writes with atomic operations via CUDA streams or sequential stages in prior work~\cite{huang2020strassen} introduces strict ordering dependencies, severely constraining parallelism.

\paragraph{Performance Modeling}
To maximize performance, it is crucial to select the optimal LCMA based on specific input shapes and hardware characteristics. 
While some existing works attempt to model LCMA performance~\cite{huang2020strassen,li2011strassengpu, huang2016strassen, gopalakrishnan2021msmes}, these analyzes typically evaluate the entire algorithm and are weak in addressing the characteristics of pipeline overlapping over computation and memory access, which are critically important in modern hardware.

Although various works focus on specific aspects of LCMA deployment, to our knowledge, no existing work simultaneously provides a unified LCMA framework with cross-platform support, optimized parallel execution, and fine-grained performance modeling. Thus, we propose \falcongemm{} to address all these issues.

\begin{figure*}[htbp]
    \centering
    \includegraphics[width=1\linewidth]{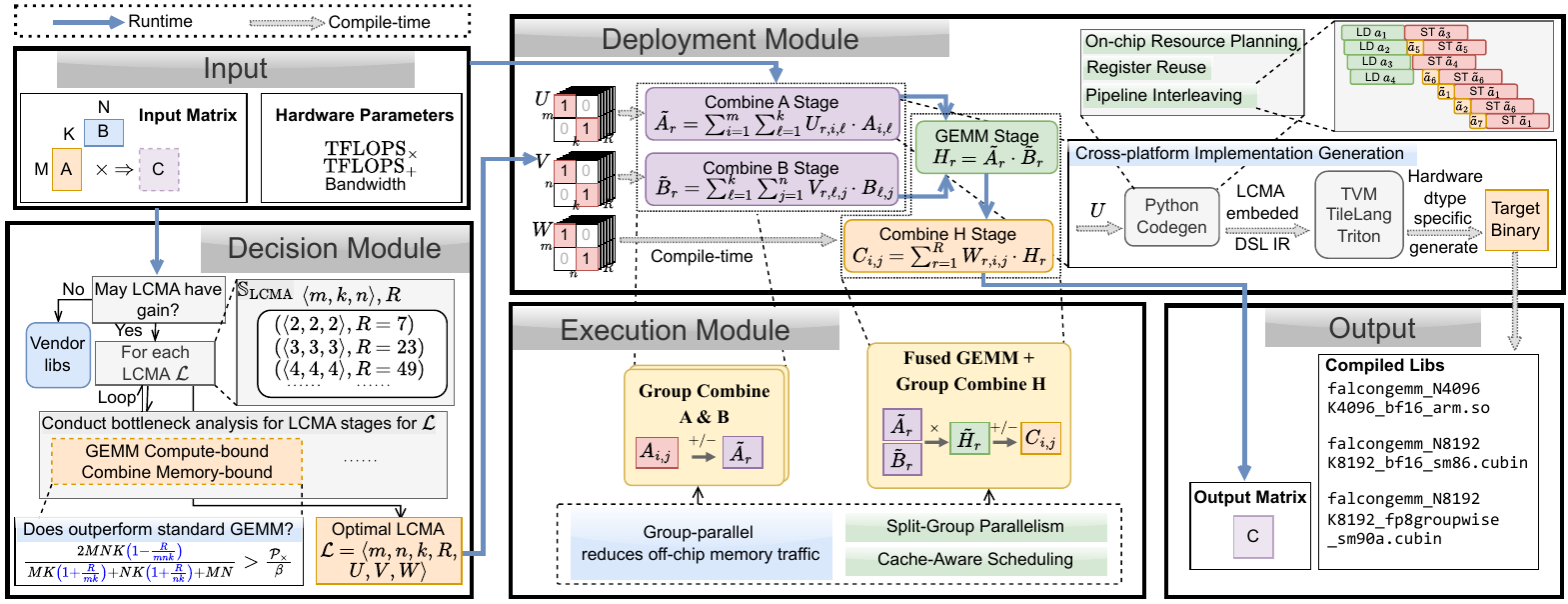}
    \caption{Overview of the workflow and internals of \falcongemm{}.}
    \label{fig:overview}
\vskip -0.15in
\end{figure*}

\section{Method}
\label{sec:method}

A cross-platform framework named \falcongemm{} is designed to generate high-performance LCMA implementations with cross-platform support. Its workflow is illustrated in Figure~\ref{fig:overview}. \falcongemm{} consists of three core components:  
\circled{1}  The Deployment Module introduces a workflow to deploy the LCMA algorithm across diverse backends (e.g., GPUs, CPUs) and configurations (e.g., data type, input shapes) using automated code generation.
\circled{2}  The Execution Module proposes a Group-Parallel Optimization to reduce the bandwidth overhead of materializing intermediate results and to avoid write conflicts.
\circled{3}  The Decision Module employs a light-weight  cost model to automatically select the performance-optimal strategy for \falcongemm{} based on the given matrix shape and hardware profile.

\subsection{Deployment Module}
\label{subsec:cross_platform}

The LCMA workflow is introduced to implement the LCMA fast and efficiently. Then automated code generation is applied to serve as a unified execution template using various LCMAs across diverse hardware architectures.

\subsubsection{LCMA Workflow}

The LCMA workflow can be decomposed into four stages, directly corresponding to the mathematical formulation in Equations~\ref{eq:addA} to ~\ref{eq:addH}. A direct implementation of LCMA is demonstrated by Algorithm~\ref{alg:simplified_workflow}. Considering the memory hierarchy for modern computing chips on both CPU and GPU,  the workflow in Algorithm~\ref{alg:simplified_workflow} operates as follows: 

\begin{description}
   \item[Combine A:]  This stage prepares the inputs $\tilde{A}_r$ for the core GEMM stage. We apply the algorithm-specific coefficient tensors $U$ to the input matrix $A$ using equation~\ref{eq:addA}. From line 2 to line 4, since $U_{r,i,\ell}$ is sparse and has its element value in $\{-1,0, 1\}$, only those matrices with non-zero coefficients are loaded into the CPU cache or the shared memory of the GPU. Finally, the accumulated results are stored back into main memory in line 5. 
   This step encodes the input matrices into LCMA's rank space.
   \item[Combine B:]   This stage works similarly to the above stage.
   \item[GEMM:]  This stage employs a batched GEMM with identical submatrix dimensions over $R$ by multiplying intermediate matrix pairs $\tilde{A}_r$ and $\tilde{B}_r$ to generate the result $H_r$ with lines 12 and 13. Note that $\tilde{A}_r$, $\tilde{B}_r$, and $H_r$ may be large enough and need to be loaded from or written to the main memory. Highly optimized vendor libraries (e.g., cuBLAS~\cite{nvidia2025cublas} or Intel MKL~\cite{intel2025mkl}) can be applied here. 
   \item[Combine H:]    Finally, the intermediate results $H_r$ are aggregated using the algorithm coefficient $W$ to transform from the algorithm's rank space back to the final output matrix $C$. To save memory bandwidth, for each $C_{i,j}$, only the $H_r$ with non-zero coefficients are needed to be loaded with lines from 16 to 17. 
\end{description}

There are three obstacles for the above algorithm to achieve both wide portability and high efficiency. 
The first obstacle is that the theoretical reduction in arithmetic operations (FLOPs) comes at a clear cost of additional memory overhead introduced by Stages 1, 2, and 4. To mask these overheads, optimized fused execution is presented in Section~\ref{sec:execution}. 
The second obstacle is how to select and determine parameters for LCMA workflow on diverse hardware configurations for an optimal implementation. This will be discussed in Section~\ref{subsec:model}.
The last obstacle is to enable performance portability with various LCMAs and hardware diversity. We fix this issue with automated code generation as below.

\subsubsection{Automated Code Generation}
We introduce an \textbf{Automated Code Generation} pipeline to decouple this engineering complexity and minimize the deployment costs with high-performance execution.

\textbf{To decouple hardware, data types, and tuning configurations},
we leverage deep learning compilers (e.g., TVM~\cite{chen2018tvm}, TileLang~\cite{wang2025tilelang}, and Triton~\cite{tillet2019triton}) to decouple the implementation logic from hardware architectures ($\mathcal{N}_{hw}$), data types ($\mathcal{N}_{dtype}$), and tuning configurations ($\mathcal{N}_{config}$). 
These compilers provide a unified Intermediate Representation (IR) that transforms abstract computation logic and memory hierarchy optimizations into highly efficient, cross-platform implementations with flexible data types and tiling configurations~\cite{chen2018tvm, feng2023tensorir, zhou2025helix}.
By specifying the target hardware architecture and shape-dependent tuning parameters during Just-In-Time (JIT) compilation, a single generic template can be automatically lowered to highly optimized machine code for arbitrary matrix dimensions and hardware platforms. This effectively neutralizes the complexity of hardware heterogeneity.

\begin{algorithm}[t]
\caption{LCMA Workflow}
\label{alg:simplified_workflow}
\small
\begin{algorithmic}[1]
\Require Inputs Matrices $A, B$;
\Require Static Algorithm Parameters $\mathcal{L} = \langle m, k, n, R, U, V, W \rangle$
\Ensure Output Matrix $C$

\State \textbf{\textcolor{blue}{//Stage 1: Combine A using Equation~\ref{eq:addA}}}
\State \textbf{For} $r \in [1, R]$: \quad \quad \quad  \quad \textcolor{blue}{//Prefetch to cache or share memory} 
\State \quad \textbf{For} $i\in[1,m], \ell \in[1,k], \text{and } U_{r,i,\ell} \neq 0$: 
    \State \quad \quad Load $A_{i,\ell}$  
    \State \quad \quad $\tilde{A}_r +=  U_{r,i,\ell} A_{i,\ell}$  \quad \textcolor{blue}{//Accumulate $\tilde{A}_r$ to main memory}

\State \textbf{\textcolor{blue}{//Stage 2: Combine B using Equation~\ref{eq:AddB}}}
\State \textbf{For} $r \in [1, R]$:  \quad \quad \quad  \quad \textcolor{blue}{//Prefetch to cache or share memory} 
\State \quad \textbf{For} $\ell\in[1,k], j \in[1,n], \text{and } V_{r,\ell,j} \neq 0$: 
    \State \quad \quad Load $B_{\ell,j}$ 
    \State \quad \quad $\tilde{B}_r += V_{r,\ell,j} B_{\ell,j}$   \quad \textcolor{blue}{//Accumulate $\tilde{B}_r$ to main memory} 

\State \textbf{\textcolor{blue}{//Stage 3: GEMM using Equation~\ref{eq:BatchedGEMM}}}
\State \textbf{For} $r \in [1, R]$:
    \State \quad $H_r = \tilde{A}_r \times \tilde{B}_r$   \quad   \quad  \quad  \textcolor{blue}{//Load AB and store H to main memory} 

\State \textbf{\textcolor{blue}{//Stage 4: Combine H using Equation~\ref{eq:addH}}}
\State \textbf{For} $i \in [1, m], j \in [1, n]$:
\State \quad \textbf{For} $r \in [1, R], \text{and } W_{r,i,j} \neq 0$:
 \textcolor{blue}{//Prefetch} 
    \State  \quad \quad  Load $H_{r}$ 
    \State \quad  \quad   $C_{i,j} += W_{r,i,j} H_r$   \quad  \textcolor{blue}{//Accumulate $C_{i,j}$ to main memory} 
\end{algorithmic}
\end{algorithm}

\textbf{To decouple various LCMAs}, the challenge lies in generating efficient IR for the diverse logic of different LCMAs ($\mathcal{N}_{algo}$). We employ a meta-programming engine to synthesize kernel code based on the specific LCMA configuration.
Given an algorithm $\mathcal{L} = \langle m, k, n, R, U, V, W \rangle$, the engine encodes the coefficient tensors directly into the IR as compile-time constants. This eliminates runtime memory access for coefficients and enables the compiler to perform constant folding, effectively pruning operations where coefficients are zero.
Moreover, three advanced micro-optimizations are applied in the IR generation procedure:  \circled{1} \textit{On-chip Resource Planning} is used to  evaluate the consumption of on-chip resources (e.g., registers and GPU shared memory) based on the LCMA coefficients. We automatically adjust the tiling strategy or decompose the computation to fit within the available resources if the estimated usage exceeds hardware limits.  \circled{2} \textit{Register Reuse} is applied by reordering arithmetic operations. This is critical for high-rank algorithms where register pressure is high and expensive off-loading spills to last-level cache or memory should be avoided.  \circled{3} \textit{Instruction Interleaving} is adopted to decompose asynchronous memory copy instructions and pipeline them with interleaved arithmetic logic. This exploits Instruction Level Parallelism (ILP) to effectively hide memory latency during the linear combination phases~\cite{goto2008toms}.

The automated code generation technique employs meta-programming/micro-optimization for specific LCMA and hardware/input aware JIT compilation in the implementation  of Algorithm~\ref{alg:simplified_workflow}.  This ensures an implementation with both wide portability and high efficiency.

\subsection{Execution Module}
\label{sec:execution}
Algorithm~\ref{alg:simplified_workflow} relies on off-chip memory to transfer intermediate data between stages, resulting in memory overhead.
It is necessary to apply fusion optimization across stages to eliminate this overhead.
As discussed in Section~\ref{sec:related}, existing fusion solutions~\cite{benson2015framework, huang2016strassen, huang2020strassen} with $R$ parallel tasks have drawbacks such as redundant memory accesses, operator fragmentation, and parallel write conflicts.
In this section, we propose the \textbf{Group-Parallel Optimization} to avoid all these issues while achieving stage fusion.
Based on this optimization, we also introduce \textbf{Split-Group Parallelism} and \textbf{Cache-Aware Scheduling} to address the new challenges of coarsened parallel granularity and cache thrashing.

\begin{figure}[t]
    \centering
    \begin{tabular}{c}
         \includegraphics[width=0.99\linewidth]{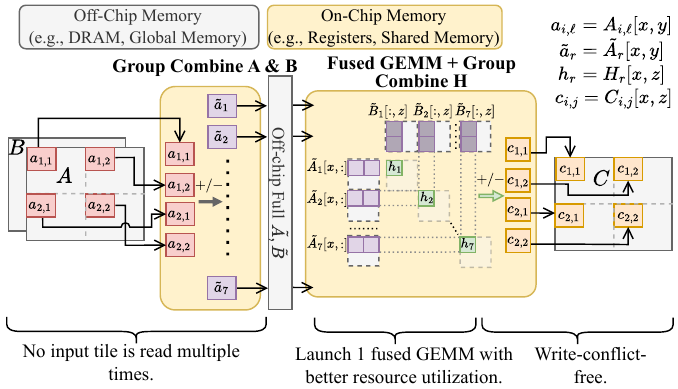} \\
    \end{tabular}
    \caption{Our proposed Group-Parallel Optimization on Strassen ($\langle 2,2,2 \rangle, R=7$) as an example.  In \textit{Group Combine A/B}, a single parallel unit processes an entire group $\{\tilde{A}_r[x,y]\}_{r=1}^{R}$ or $\{\tilde{B}_r[y,z]\}_{r=1}^{R}$ by transforming 4 pairs of input elements into a group of 7 pairs of intermediate elements entirely on-chip. \textit{Fused GEMM + Group Combine H} transforms the group $\{H_r[x,z]\}_{r=1}^{R}$  into $\{C_{i,j}[x,z]\}_{i=1\ j=1}^{m\ \ \ n}$ directly without write conflicts.}
    \label{fig:thiswork}
\end{figure}

\begin{algorithm}[t]
\caption{Fused LCMA Workflow}
\label{alg:fused_combineA}
\small
\begin{algorithmic}[1]
\Require Inputs Matrices $A, B$;

\Ensure Output Matrix $C$

\State \textbf{\textcolor{blue}{//Stage 1: Group Combine A}}
\State \textbf{Parallel For} $x \in [1, \lceil \frac{M}{m} \rceil], y \in [1, \lceil \frac{K}{k} \rceil]$:
    \State \quad \textit{\textcolor{blue}{// Load inputs to on-chip memory and compute combinations}}
    \State \quad \textbf{For} $i \in [1, m], \ell \in [1, k]$:
        \State \quad \quad \textbf{Load} $A_{i,\ell}[x,y]$ from off-chip memory
    
    \State \quad \textit{\textcolor{blue}{// Process all elements in group $\{\tilde{A}_r[x,y]\}_{r=1}^{R}$}}
    \State \quad \textbf{For} $r \in [1, R]$:
        \State \quad \quad \textbf{For} $i \in [1, m], \ell \in [1, k], \text{and }U_{r,i,\ell} \neq 0$:
            \State \quad \quad \quad $\tilde{A}_r[x,y] += U_{r,i,\ell} \cdot A_{i,\ell}[x,y]$ 

\State \textbf{\textcolor{blue}{//Stage 2: Group Combine B}}
\State \textbf{Parallel For} $y \in [1, \lceil \frac{K}{k} \rceil], z \in [1, \lceil \frac{N}{n} \rceil]$:
    \State \quad \textit{\textcolor{blue}{// Load inputs to on-chip memory and compute combinations}}
    \State \quad \textbf{For} $\ell \in [1, k], j \in [1, n]$:
        \State \quad \quad \textbf{Load} $B_{\ell,j}[y,z]$ from off-chip memory
    
    \State \quad \textit{\textcolor{blue}{// Process all elements in group $\{\tilde{B}_r[y,z]\}_{r=1}^{R}$}}
    \State \quad \textbf{For} $r \in [1, R]$:
        \State \quad \quad \textbf{For} $\ell \in [1, k], j \in [1, n], \text{and }V_{r,\ell,j} \neq 0$:
            \State \quad \quad \quad $\tilde{B}_r[y,z] += V_{r,\ell,j} \cdot B_{\ell,j}[y,z]$ 

\State \textbf{\textcolor{blue}{//Stage 3/4: Fused GEMM and Group Combine H}}
\State \textbf{Parallel For} $x \in [1, \lceil \frac{M}{m} \rceil], z \in [1, \lceil \frac{N}{n} \rceil]$:
    \State \quad \textit{\textcolor{blue}{// Process all elements in group $\{H_r[x,z]\}_{r=1}^{R}$}}
    \State \quad \textbf{For} $r \in [1, R]$: 
        \State \quad \quad \textit{\textcolor{blue}{// Accumulate $r$ from 1 to R on-chip (Register/Shared Mem)}}

        \State \quad \quad \textbf{For} $y \in [1, \lceil \frac{K}{k} \rceil]$:
            \State \quad \quad \quad $H_r[x,z]  += \tilde{A}_r[x, y] \times \tilde{B}_r[y, z]$
        
        \State \quad \quad \textit{\textcolor{blue}{// On-chip update output C, without any write conflicts}}
        \State \quad \quad \textbf{For} $i \in [1, m], j \in [1, n], \text{and } W_{r,i,j} \neq 0$:
            \State \quad \quad \quad $C_{i,j}[x, z] +=  W_{r,i,j} \cdot H_r[x,z]$
            
    \State \quad \textbf{Store} $\{C_{i,j}[x,z]\}$ to off-chip memory
\end{algorithmic}
\end{algorithm}

\paragraph{Group-Parallel Optimization} 
A \textbf{Group} is the collection of elements across the $R$ dimension at the same relative coordinate. 
Specifically, let $(x,y)$, $(y,z)$, and $(x,z)$ denote the coordinates of the elements (or blocks) $A_{i,\ell}[x,y]$ in $A_{i,\ell}$, $B_{\ell,j}[y,z]$ in $B_{\ell,j}$ and $C_{i,j}[x,z]$ in $C_{i,j}$. Here $x \in [1, \lceil \frac{M}{m} \rceil]$, $y \in [1, \lceil \frac{K}{k} \rceil]$ and $z \in [1, \lceil \frac{N}{n} \rceil]$, the group in Combine A is defined as $\{\tilde{A}_r[x,y]\}_{r=1}^{R}$, similarly group in Combine B is $\{\tilde{B}_r[y,z]\}_{r=1}^{R}$. The group used in GEMM and Combine H is defined as $\{H_r[x,z]\}_{r=1}^{R}$.

The \textbf{Group-Parallel Optimization} utilizes the inherent data locality in a group. That means all elements within a group can be computed by the same parallel execution unit. 
For example, in the Combine A stage, computing the elements at a specific relative coordinate $(x,y)$ consisting of all $R$ $\tilde{A}_r$ matrices (i.e., $\{\tilde{A}_r[x,y]\}_{r=1}^{R}$)  requires accessing \textit{only} the source elements at the same $(x,y)$ position within the $m \times k$ input submatrices $A_{i, \ell}$ (i.e., $\{A_{i,\ell}[x,y]\}_{i=1\ \ell=1}^{m\ \ \ k}$). 
Crucially, these source elements contribute \textit{exclusively} to this specific group, i.e., no $A_{i,\ell}[x,y]$ contributes to computing $\tilde{A}_r[x',y']$ where $x\neq x'$ or $y \neq y'$. By loading $A_{i,\ell}[x,y]$ exactly once, a parallel unit can exclusively compute all its contributing $\tilde{A}_r[x,y]$ elements in the group. This special data locality is illustrated in Figure~\ref{fig:thiswork}. 

This feature is not limited to Combine A but applies to all other combine stages in Algorithm~\ref{alg:simplified_workflow}. The linear combination to obtain a group of $\{\tilde{B}_r[y,z]\}_{r=1}^{R}$ from $\{B_{\ell,j}[y,z]\}_{\ell=1 j=1}^{k\ \ \ n}$ shares the exact same memory footprint locality as in the Combine A stage. 
For the GEMM and Combine H stages illustrated in Figure~\ref{fig:thiswork}, once the elements in group $\{H_r[x,z]\}_{r=1}^{R}$ are computed locally, the output elements $\{C_{i,j}[x,z]\}_{i=1\ j=1}^{m\ \ \ n}$ can be intuitively derived from them without external dependencies.
Note our design choice to materialize $\{\tilde{A}_r\}_{r=1}^{R}$ and $\{\tilde{B}_r\}_{r=1}^{R}$ to off-chip memory, as fully fusing all stages without introducing extra recomputation and memory access would demand infeasibly large on-chip resources.

To implement this optimization, loop reordering is applied to enable Group-Parallel Optimization in Algorithm~\ref{alg:fused_combineA}. We adjust the loop order to keep $(x,y)$ and $(y,z)$ as the outermost loops, and $(m,k)$ and $(k,n)$ as the innermost loops for the first two combine stages using codes from line 2 to line 9, and codes from line 11 to line 18 respectively. After this optimization, one can see that Algorithm~\ref{alg:simplified_workflow} accesses matrix $A$ for $||U_{r,i,\ell}||_{0} \times \frac{M}{m} \times \frac{K}{k}$ times, and matrices $U_{r,i,\ell}$ and $\tilde{A}_r$ once each, here $||U_{r,i,\ell}||_{0}$ records the number of non-zero elements in $U_{r,i,\ell}$. Whereas Algorithm~\ref{alg:fused_combineA} reads $A$ and writes $\tilde{A}_r$ only once, at the cost of reading matrix $U_{r,i,\ell}$ for $\frac{M}{m} \times \frac{K}{k}$ times. We are actually making a tradeoff by saving memory accesses on $A$ at the cost of $U_{r,i,\ell}$. Note that matrix $U_{r,i,\ell}$ is a sparse matrix with few non-zero elements. We can store it as an adjacency matrix in an array to reduce its size, and then store it directly in the last-level D-cache in CPU or shared memory in GPU to avoid main memory accesses. But in this work, we apply code generation techniques and compile the coefficients of $U$ into code and store it in the I-Cache. Finally, we eliminate all the extra data accesses on Combine A and also B similarly.  

Applying group-parallel can also achieve an efficient fusion of GEMM and Combine H stages with data locality, as shown in Algorithm~\ref{alg:fused_combineA}. 
We first compute the group elements $\{H_r[x,z]\}_{r=1}^{R}$ by multiplying the matrices $\tilde{A}_r$ and $\tilde{B}_r$ by looping over the $y$ dimension in line 24. Then the set of output elements $\{C_{i,j}[x,z]\}_{i=1\ j=1}^{m\ \ \ n}$ is calculated \textit{exclusively} on group $\{H_r[x,z]\}_{r=1}^{R}$, which is at the same coordinate in line 28. Thus a single computation unit can possess all elements in group $\{H_r[x,z]\}_{r=1}^R$ only once, and accumulate them into the on-chip element $C_{i,j}$. The final results are then written directly to the output matrix. This minimizes memory traffic without incurring extra computation or write conflicts.

\paragraph{Split-Group Parallelism} 
Group-Parallel Optimization using the group as a coarser parallel granularity may lead to a load imbalance problem.  As a single compute unit must compute an entire group containing $R$ elements for the Fused GEMM and Combine H stages, for GPU devices, this coarse scheduling granularity can lead to resource waste. 
For example, scheduling $4$ groups on $3$ Streaming Multiprocessors (SMs) leaves one SM processing the last group alone while the others are idle (Figure~\ref{fig:fusion_evolution}(a)). 
In practice, when a matrix multiplication with $(M,N,K) = (4096,4096,4096)$ is computed with $128 \times 128$ GEMM tile block on $M,N$ dimension for the NVIDIA H20 GPU ($78$ SMs) using Strassen's algorithm ($R=7$). The total number of $H_r$ tile blocks is $\frac{4096/2}{128} \times \frac{4096/2}{128} \times 7 = 1792$, or equivalent to $256$ groups. If processing one $H_r$ tile takes one wave, scheduling at the tile block level yields $\lceil 1792/78 \rceil = 23$ waves per SM. In contrast, scheduling at the group level yields $\lceil 256/78 \rceil = 4$ groups ($4 \times 7 = 28$ waves) per SM, which incurs about $\frac{28-23}{23}\times 100\%=21.7\%$ waves or running time waste.

\begin{figure}[t]
    \small
    \centering
    \newlength{\colwidth}
    \setlength{\colwidth}{0.145\textwidth} 
    
    \begin{tabular}{ccc}
        \includegraphics[width=\colwidth]{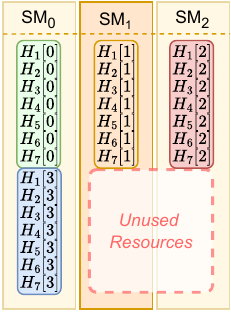} &
        \includegraphics[width=\colwidth]{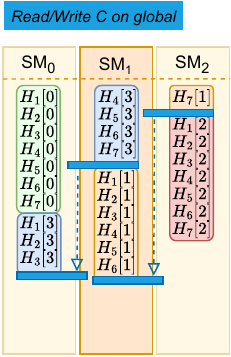} &
        \includegraphics[width=\colwidth]{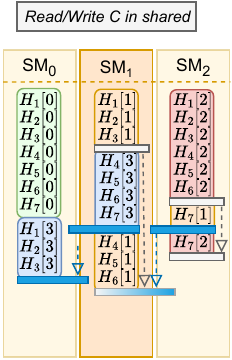} \\
        (a) Group-Parallel & (b) Split-Group & (c) Cache-Aware \\
    \end{tabular}
    \caption{Evolution of the \falcongemm{} Group-Parallel Optimization on GPU, in total of 4 different coordinates $(x,z)$, denoted as $H_r[0]\sim H_r[3]$. (a) \textit{Group-Parallel} assigns group to CTAs to SMs, leading to resource waste. (b) \textit{Split-Group} uses persistent kernels to achieve fine-grained scheduling. (c) \textit{Cache-Aware} scheduling reorders execution within each SMs to prevent cache thrashing.}
    \label{fig:fusion_evolution}
\end{figure}

To resolve this issue, \textbf{Split-Group Parallelism} adopts persistent kernel techniques~\cite{perks} to support tile block level scheduling and enable fine-grained granularity parallelism without breaking the data locality feature.  In Figure~\ref{fig:fusion_evolution}(b), the persistent kernel techniques explicitly control tile scheduling over the lifecycle of each SM by keeping a single Cooperative Thread Array (CTA) alive throughout the entire GEMM execution. 
Instead of treating groups as indivisible units, 
persistent kernel allows a single group to be split across at most two SMs, effectively restoring fine-grained execution efficiency. 
Specifically, we calculate a precise tile capacity per SM to distribute the total tiles evenly. When assigning a group to an SM exceeds the remaining capacity, we split the group: the current SM fills its capacity with the initial tiles of the group, and the overflow portion is processed by the next SM. Crucially, atomic operations are not required even when a group is split across two SMs. The deterministic execution order ensures that one SM processes its portion of the group in its earliest waves, while the other processes the remainder in its final waves, naturally avoiding write conflicts to the output.

\paragraph{Cache-Aware Scheduling} 
On some hardware, we observe that cache thrashing can become a new performance bottleneck. In the previous example of matrix multiplication with $(M,N,K) = (4096,4096,4096)$ on an H20 GPU, Split-Group scheduling reduces the number of execution waves, but the heavy memory burden from a low L2 cache hit rate triggers the GPU power limit, dropping the SM frequency from 1.80 GHz to 1.61 GHz and increasing the overall time. This low L2 cache hit rate occurs because concurrent SMs access data stored far apart in memory. 
For example, concurrent SMs process $H_1[0]$, $H_4[3]$, and $H_7[1]$ during the initial wave in Figure~\ref{fig:fusion_evolution}(b), which requires accessing data in $\tilde{A}_1$, $\tilde{A}_4$, and $\tilde{A}_7$. However, $\tilde{A}_1$, $\tilde{A}_4$, and $\tilde{A}_7$ are stored far apart, as elements with the same $r$ are stored contiguously. 
This scattered access pattern severely degrades the L2 cache hit rate.

To mitigate this, we propose a \textbf{Cache-Aware Scheduling} policy that reorders the computation sequence of $H_r$ across groups within each SM. After reordering, different SMs access data corresponding to the same $r$ during the majority of execution waves. As shown in Figure~\ref{fig:fusion_evolution}(c), concurrent SMs now process data with the same $r$ like $H_1[0]$, $H_1[1]$, and $H_1[2]$, all of which require elements in $\tilde{A}_1$. Because this reordering interleaves the computation of different groups, their accumulators need to be buffered using extra on-chip resources. However, this trade-off yields significantly improved memory system efficiency and overall performance.

\subsection{Decision Module}
\label{subsec:model}

Given an input matrix shape $(M, N, K)$ and a specific hardware platform, the \textbf{Decision Module} iterates over a candidate set of LCMAs $\mathbb{S}_{LCMA}$ to identify the optimal LCMA or defaults to standard GEMM if no LCMA is likely to yield a gain. In this section, we will first abstract hardware parameters and then summarize our LCMAs with a robust theoretical arithmetic intensity analysis.

Selecting the optimal algorithms requires balancing the reduction in arithmetic complexity against the overhead of linear combination steps. This trade-off is strictly governed by the input matrix dimensions and the specific performance characteristics of the hardware platform. 
Thus a robust theoretical model is essential to determine whether an LCMA yields a performance gain and to identify which specific algorithm maximizes this benefit.

We first abstract the hardware platform with the tuple of ($\text{FLOPS}_{\times}, \text{FLOPS}_{+}, \beta$) and then conduct an arithmetic intensity analysis on standard GEMM for a better understanding. 
In our hardware-aware three tuple ($\text{FLOPS}_{\times}, \text{FLOPS}_{+}, \beta$), $\text{FLOPS}_{\times}$ is the floating-point operations per second for matrix multiplications in the GEMM stage
(e.g. Tensor Cores or SME instructions),
$\text{FLOPS}_{+}$ is defined as the floating-point operations per second  for addition or subtraction in Combine A/B/H stages 
(e.g., CUDA Cores or SVE instructions), and $\beta$ is defined as the bandwidth of off-chip memory for a target data type. 
When a standard GEMM is bounded by memory with 
\begin{equation}
\label{eq:stdgemm}
\small
\frac{2MNK}{MK + NK + MN} \leq \frac{\text{FLOPS}_{\times}}{\beta}
\end{equation}
then all LCMAs are unlikely to yield a gain, as they reduce computation at the cost of increasing total memory traffic. 
In such cases, we immediately return the standard GEMM.
Otherwise, it is compute-bound, and we take
$\frac{2MNK}{\text{FLOPS}_{\times}}$ as its running time.

\begin{table}[b]
\centering
\caption{Theoretical arithmetic intensity analysis of LCMA Stages. 
$\|U\|_0, \|V\|_0, \|W\|_0$ represent the number of non-zero elements in the coefficient tensors. 
The Arithmetic Intensity is the ratio of Computation to Memory Access.}
\label{tbl:cost_model}
\renewcommand{\arraystretch}{1.5} % Increase row height for better readability
\setlength{\tabcolsep}{3pt}       % Reduce column padding to fit width
\resizebox{\columnwidth}{!}{%     % Resize to fit single column
\begin{tabular}{|c|c|c|c|}
\hline
\textbf{Stage} & \textbf{\makecell{Computation \\ (FLOPs)}} & \textbf{\makecell{Memory Access \\ (Number of elements)}} & \textbf{\makecell{Arithmetic Intensity}} \\ \hline \hline

\makecell{Standard GEMM} & 
$2MNK$ & 
\makecell{$MK + NK + MN$} & 
$\frac{2MNK}{MK + NK + MN}$ \\ \hline

\makecell{Combine A Stage \\ (Algo.~\ref{alg:simplified_workflow} Stage 1)} & 
$(\|U\|_0 - R)\cdot \frac{M}{m}\cdot \frac{K}{k}$ & 
\makecell{$MK(1+\frac{R}{mk})$} & 
$\frac{\|U\|_0 - R}{mk + R}$ \\ \hline

\makecell{Combine B Stage\\ (Algo.~\ref{alg:simplified_workflow} Stage 2)} & 
$(\|V\|_0 - R)\cdot \frac{N}{n}\cdot \frac{K}{k}$ & 
\makecell{$NK(1+\frac{R}{nk})$} & 
$\frac{\|V\|_0 - R}{nk + R}$ \\ \hline

\makecell{GEMM Stage\\ (Algo.~\ref{alg:simplified_workflow} Stage 3)} & 
$\frac{2R MNK}{mnk}$ & 
\makecell{$R(\frac{MK}{mk} + \frac{NK}{nk} + \frac{MN}{mn})$} & 
$ \frac{2 MNK}{nMK + mNK+ kMN}$ \\ \hline

\makecell{Combine H Stage\\ (Algo.~\ref{alg:simplified_workflow} Stage 4)} & 
$(\|W\|_0 - mn)\cdot \frac{M}{m}\cdot \frac{N}{n}$ & 
\makecell{$MN(1+\frac{R}{mn})$} & 
$\frac{\|W\|_0 - mn}{R + mn}$ \\ \hline

\end{tabular}%
}
\end{table}

Now we conduct an arithmetic intensity analysis on the LCMA described as $\mathcal{L} = \langle m,k,n, R, U, V, W \rangle$ and then summarize our robust theoretical model. Combine A stage in Algorithm~\ref{alg:fused_combineA}
involves addition/subtraction on $m \times k$ submatrices, and each submatrix contributes $\frac{M}{m}\cdot \frac{K}{k}$ addition/subtraction operations. To obtain $R$ matrices $\tilde{A}_r$, a total of $(\|U\|_0 - R)$ submatrix additions/subtractions are needed (for example, in Strassen's algorithm, $\|U\|_0 = 12$ and $R = 7$, so beside $R=7$ matrix assignment, only $5$ submatrix additions/subtractions are performed). Hence, the total number of addition/subtraction operations in Combine A is $(\|U\|_0 - R)\cdot \frac{M}{m}\cdot \frac{K}{k}$.
Regarding memory access, \falcongemm{} requires loading the $A$ matrix only once ($MK$ elements) and outputs $R$ $\tilde{A}_r$ matrices ($R\cdot \frac{M}{m}\cdot \frac{K}{k}$ elements), thus totally there are $MK(1+\frac{R}{mk})$ elements.
If the $\frac{(\|U\|_0 - R)\cdot \frac{M}{m}\cdot \frac{K}{k}}{MK(1+\frac{R}{mk})} = \frac{\|U\|_0 - R}{mk + R}> \frac{\text{FLOPS}_{+}}{\beta}$ is satisfied, then 
the estimated running time is $\frac{(\|U\|_0 - R)\cdot \frac{M}{m}\cdot \frac{K}{k}}{ \text{FLOPS}_{+}}$; otherwise, it is memory-bound, and the time is $\frac{MK\left(1+\frac{R}{mk}\right)}{\beta}$. Similarly, we derive and collect all computation operations, memory access, and arithmetic intensity equations in Table~\ref{tbl:cost_model} for the other three stages in an LCMA $\mathcal{L} = \langle m, k, n, R, U, V, W \rangle$. Among all the four stages, the three combine stages are likely to be memory-bound for most LCMA.

LCMA is used if and only if it can deliver a speedup over the standard GEMM.  In the most common scenario, the Combine stages are memory-bound while the GEMM stage is compute-bound. Thus, according to the formulas in Table~\ref{tbl:cost_model}, the time usage of LCMA needs to be less than the time usage of standard GEMM, that means
$
\frac{MK\left(1+\frac{R}{mk}\right)}{\beta} + \frac{NK\left(1+\frac{R}{nk}\right)}{\beta} +
\frac{2R \frac{MNK}{mnk}}{\text{FLOPS}_{\times}} + \frac{MN\left(1+\frac{R}{mn}\right)}{\beta} < \frac{2MNK}{\text{FLOPS}_{\times}}
$
It can be simplified as: 
\begin{equation}
\label{eq:lcma_condition}
\small
\frac{2MNK {\left(1 - \frac{R}{mnk}\right)}}{MK{\left(1+\frac{R}{mk}\right)} + NK{\left(1+\frac{R}{nk}\right)} + MN{\left(1+\frac{R}{mn}\right)}} > \frac{\text{FLOPS}_{\times}}{\beta}
\end{equation}

The inequality in Eq.~\ref{eq:lcma_condition} shares an interesting structural similarity with Eq.~\ref{eq:stdgemm}: it is a standard GEMM arithmetic intensity formula with extra LCMA-specific scaling factors.
It intuitively quantifies how LCMA trades off lower bandwidth for reduced arithmetic complexity.
Here, $(1 - \frac{R}{mnk})$ represents the scale of the computational savings of the GEMM, while the denominator represents the memory traffic overhead, i.e., $\frac{R}{mk}$ corresponds to the extra bandwidth for writing $\tilde{A}$ in the Combine A stage, $\frac{R}{nk}$ for writing $\tilde{B}$ in the Combine B stage, and $\frac{R}{mn}$ for writing $H$ in the Combine H stage. Note that, the optimization with fused GEMM and Combine H stage in Algorithm~\ref{alg:fused_combineA} avoids off-chip writing and reading of the intermediate matrices $H_r$, thus the overhead $\frac{R}{mn}$ associated with the $MN$ term is eliminated. Finally, Eq.~\ref{eq:lcma_condition} can be simplified as:
\begin{equation}
\label{eq:fused_condition}
\small
    \frac{2MNK \left(1 - \frac{R}{mnk}\right)}{MK{\left(1+\frac{R}{mk}\right)} + NK{\left(1+\frac{R}{nk}\right)} + MN} > \frac{\text{FLOPS}_{\times}}{\beta}
\end{equation}

In conclusion, Decision Module provides a theoretical prediction of the acceleration for each LCMA and standard GEMM algorithm. The LCMA with the highest acceleration ratio will be adopted as our final implementation, If no performance gains can be achieved with an LCMA, we fall back to the standard GEMM to keep its best performance. For limited space, extended analyses on tiling blocks and boundary scenarios are omitted for brevity.

\begin{figure*}[t]
    \centering
    \begin{tabular}{cccccc}
         \includegraphics[width=0.3\linewidth]{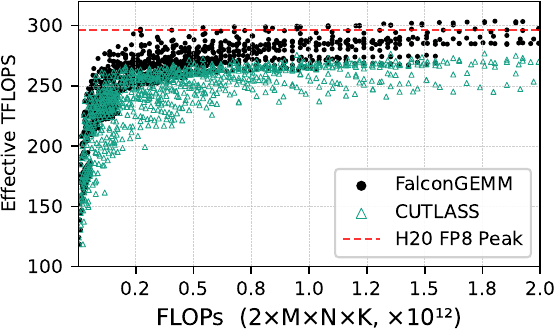}& 
         \includegraphics[width=0.3\linewidth]{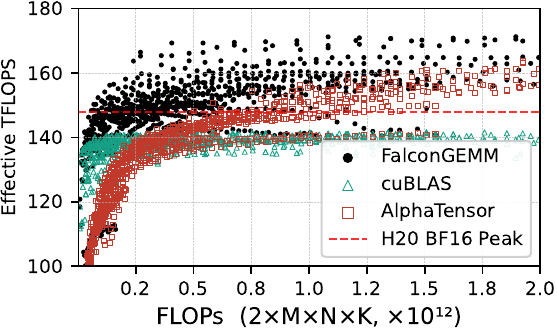} &  
         \includegraphics[width=0.3\linewidth]{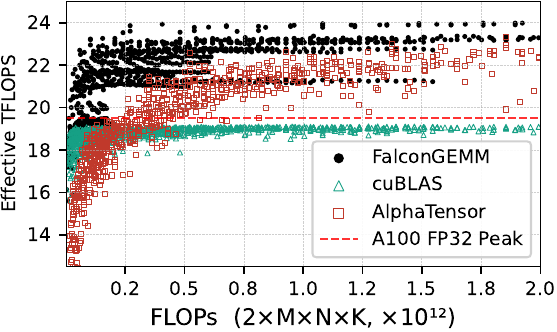} \\
         (a) NVIDIA H20 FP8 & (b) NVIDIA H20 BF16  & (c) NVIDIA A100 FP32 \\
         \includegraphics[width=0.3\linewidth]{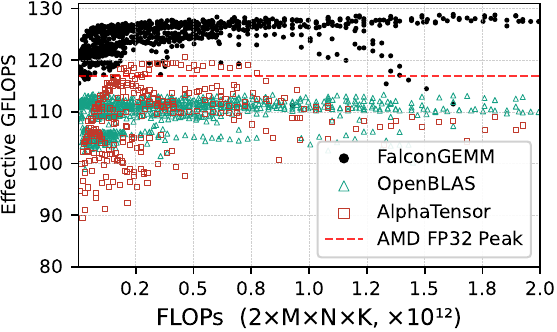} &
         \includegraphics[width=0.3\linewidth]{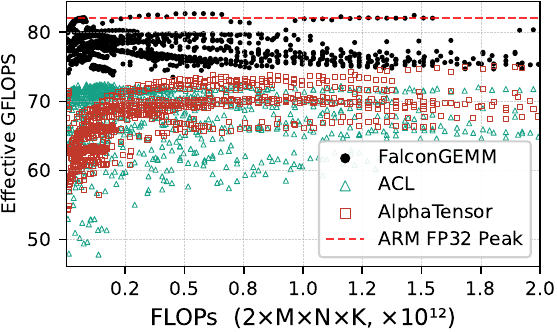} &
         \includegraphics[width=0.3\linewidth]{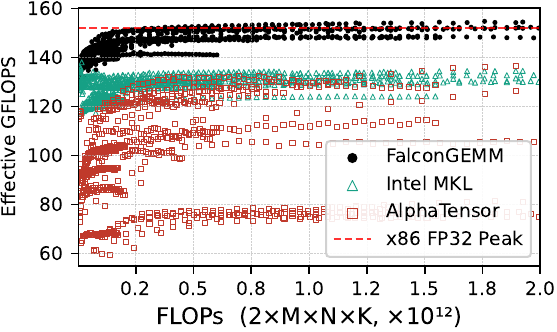} \\
         (d) AMD EPYC 9K84 FP32 & (e) ARM Neoverse-V1 FP32 & (f) Intel Xeon 8255C FP32  \\
    \end{tabular}
    \caption{Operator-level performance is compared between \falcongemm{} and six other libraries on  three LLM workloads. The x-axis represents the amount of floating-point operations calculated as $2\times M \times N \times  K$ using 960 test shapes. The y-axis is the  performance measured in effective TFLOPS or GFLOPS. The peak-breaking performance is achieved by algorithmic  operation reduction.}
    \label{fig:main_perf}
\vskip -0.3in
\end{figure*}

\section{Evaluation}
\label{sec:evaluation}

We implement \falcongemm{} with all the above optimization to demonstrate its superior efficiency in accelerating Large Language Model (LLM) workloads across diverse hardware  including both GPU and CPU devices supporting different data types (e.g., FP32, BF16, FP16 and FP8) supported by Deployment Module (Section~\ref{subsec:cross_platform}).
We compare \falcongemm{} against multiple state-of-the-art  matrix multiplication libraries and specialized LCMA implementations. 
We use step-wise evaluation to verify the performance gain of our proposed Execution Module (Section~\ref{sec:execution}), and the roofline results also confirm the selection of optimal LCMA and GEMM algorithms in our Decision Module (Section~\ref{subsec:model}).

\subsection{Experiment Setup}
\textbf{Hardware Platforms.} Our evaluation spans five representative computing devices with varying compute and memory characteristics: \circled{1} \textbf{NVIDIA H20} (Hopper, 96GB), featuring 4.0 TB/s HBM3 bandwidth, with CUDA 12.9;  \circled{2} \textbf{NVIDIA A100} (Ampere, 40GB), providing 1.6 TB/s HBM2 bandwidth, with CUDA 12.8;   \circled{3} \textbf{Intel Xeon Platinum} 8255C processor (2.50 GHz), with 240 GB/s memory bandwidth;  \circled{4} \textbf{AMD EPYC} 9K84 Processor (2.60 GHz), with 250 GB/s memory bandwidth;  \circled{5} Amazon EC2 M7g Instances using \textbf{ARM Neoverse-V1} (2.1 GHz), with 20.8 GB/s memory bandwidth.

\textbf{Workloads and Metrics.} 
We extract linear layer shapes ($N, K$) from three open-source LLMs (\textbf{DeepSeek-R1}~\cite{deepseekai2025r1}, \textbf{Qwen3.5-397B}~\cite{yang2025qwen3}, and \textbf{HunyuanVideo}~\cite{kong2024hunyuanvideo}). Supported data types across all platforms include FP32, BF16, FP16, and FP8, depending on hardware capabilities. 
For FP8, we adopt the full BF16-to-quantized-FP8 workflow with $1 \times 128$ block-wise scaling for FP8E4M3~\cite{micikevicius2022fp8}, consistent with widely used CUTLASS~\cite{nvidia2025cutlass} and DeepGEMM~\cite{deepseek2025deepgemm}.

\textbf{Baselines.} We select the state-of-the-art competitors as baselines. For standard GEMM, we use \textbf{cuBLAS}~\cite{nvidia2025cublas} (embedded in CUDA Toolkit) on NVIDIA GPUs, \textbf{Intel MKL}~\cite{intel2025mkl} on Intel x86, \textbf{OpenBLAS}~\cite{wang2013openblas} on AMD x86, and \textbf{ACL}~\cite{arm2025acl} (v52.8) on ARM. \textbf{AlphaTensor}~\cite{fawzi2022discovering} (relying on JAX) serves as the state-of-the-art cross-platform LCMA competitor. For FP8E4M3 with block-wise scaling on Hopper, we compare against  \textbf{CUTLASS} (v4.4.2)~\cite{nvidia2025cutlass}, and we find no prior LCMA competitor that supports FP8 quantization.

\textbf{LCMA Settings.}
The LCMA library of \falcongemm{} employs various LCMAs with coefficient matrices provided by the AlphaTensor codebase~\cite{alphatensorcode}, and we use $m,n,k \in [2,5]$.
Our software stack includes TileLang v0.1.8~\cite{wang2025tilelang}, Triton v3.6.0~\cite{tillet2019triton}, and TVM v0.23.0~\cite{chen2018tvm}.
The code of \falcongemm{} will be open-sourced upon publication.

\textbf{Measurement.} Performance is measured in Effective TFLOPS (or GFLOPS), defined as $\frac{2MNK}{\text{time\_seconds}} \times 10^{-12}$ (or $\times 10^{-9}$ for GFLOPS),
where $2MNK$ is the amount of floating-point operations of standard GEMM rather than LCMAs, thereby enabling a fair comparison between the LCMA and standard GEMM. 
Thus, the performance in the following experiments may \textbf{exceed the hardware's peak TFLOPS}, which is due to the algorithmic advantage of LCMA rather than  any hardware tricks like overclocking.

\subsection{Operator-level Performance Evaluation}
\label{subsec:operator_throughput}
The evaluation on operator-level performance uses 960  linear layer shapes selected from 3 LLMs. We select 11 shapes on $(N, K)$ from DeepSeek-R1~\cite{deepseekai2025r1}, 7 shapes from  Qwen3.5-397B~\cite{yang2025qwen3}, and 6 shapes from HunyuanVideo~\cite{kong2024hunyuanvideo}. The dimension $M$ increases from  512 to 20480 with a step of 512, and finally a total of $960$ combinations of $(M,N,K)$ are generated as the  test shape cases.
Figure~\ref{fig:main_perf} plots the performance results of \falcongemm{} compared with 4 other vendor-provided libraries and Deepmind's AlphaTensor~\cite{fawzi2022discovering} on these 960 test shapes. The results confirm that \falcongemm{} delivers a strong cross-hardware and data-type adaptability, and consistently outperforms standard GEMM libraries (CUTLASS, cuBLAS, MKL, OpenBLAS, ACL) and LCMA competitor AlphaTensor. \falcongemm{} achieves 7.59\%, 7.50\%,  11.82\%, 12.17\%, 17.85\%, 12.94\% higher performance than these best GEMM libraries on H20 using FP8, H20 using BF16, A100 using FP32, AMD CPU using FP32,  ARM CPU using FP32, and Intel CPU using FP32, respectively. That is because LCMAs uses fewer computations, and \falcongemm{} frequently surpasses the theoretical hardware peak which is the strict upper bound for standard GEMMs. 
The above statistics include cases where \falcongemm{} falls back to standard GEMM. If we count only cases where \falcongemm{} selects LCMA, the speedups over the best GEMM libraries become 11.20\%, 11.06\%, 17.03\%, 12.55\%, 17.85\%, and 14.33\%, respectively.
For the AlphaTensor, the only competitor on LCMA, ours surpass it with 20.47\%, 12.41\%, 17.01\%, 25.35\%, 55.61\% higher performance on all five devices besides H20 using FP8 which is not supported by AlphaTensor. The experiments also show that AlphaTensor can exceed the hardware peak on a few significantly large shapes, but for the small shapes, AlphaTensor  often  performs worse than standard GEMM. This limitation is due to the inherent overheads of traditional LCMA deployments, as we have discussed in Section~\ref{sec:related}, which make previous LCMA impractical for real-world deployments. In contrast, \falcongemm{} achieves peak-breaking efficiency even on small matrices by eliminating these overheads, proving its robust practicality for real-world LLM inference.

\begin{figure}[t]
    \centering
    \includegraphics[width=0.9\linewidth]{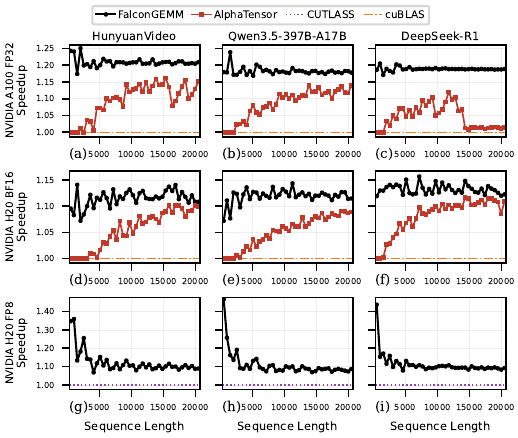}
    \caption{End-to-end LLM speedup on PyTorch using different GEMM backends on H20 and A100 using FP32, BF16 and FP8. Here the x-axis is the sequence length (M) range from 128 to 20K, the y-axis is the relative speedup compared to PyTorch baselines using CUTLASS or cuBLAS. 
            }
    \label{fig:llm_speedup}

\end{figure}

\subsection{End-to-End LLM Performance}
\label{subsec:e2e_performance}

The end-to-end (e2e) LLM performance evaluation on HunyuanVideo, Qwen3.5 and DeepSeek-R1 is conducted by replacing the GEMM operations of the linear layers in PyTorch with  \falcongemm{} and AlphaTensor. For \falcongemm{}, we use offline Combine B stage for static weights $B$. 
The e2e performance results for the prefill stage on GPUs across varying sequence lengths are recorded and illustrated in Figure~\ref{fig:llm_speedup}.
PyTorch using \falcongemm{} delivers a highly stable and significant performance gain across all sequence lengths. For example, compared with PyTorch, $18.12\%, 12.24\%$, and $11.46\%$ averaged e2e performance gains in prefill are achieved on the A100 using FP32, H20 using BF16 and H20 using FP8, respectively.  This stability is attributed to our Decision Module, which intelligently assigns the optimal LCMA configuration for specific weight shapes even when $M$ is small. 
For example, PyTorch has recorded that $97.9\%, 85.7\%$, and $ 57.7\%$ linear layers use \falcongemm{}'s LCMA on HunyuanVideo, Qwen3.5-397B, and DeepSeek-R1 models on H20 using FP8.
This is the lowest ratio among six hardware and datatype cases.  In contrast, prior LCMA frameworks like AlphaTensor fail to provide consistent benefits. They typically yield no speedup at small sequence lengths due to unoptimized overheads. More critically, at very large sequence lengths (e.g., on A100 using FP32), AlphaTensor's performance degrades sharply. We observe that their large-$R$ algorithms consume excessive shared memory, forcing the system to fall back to the basic Strassen's algorithm. 
\falcongemm{}'s execution entirely overcomes these limitations, ensuring robust acceleration across all deployment scenarios.
For FP8 workloads with small $M$, where quantization overhead is typically substantial, \falcongemm{} fuses the quantization into the Combine A stage, yielding at most $46\%$ greater performance.

\subsection{Step-wise Evaluation for Execution Module}
\label{subsec:ablation}

\begin{figure}[t]
    \centering
    \includegraphics[width=0.9\linewidth]{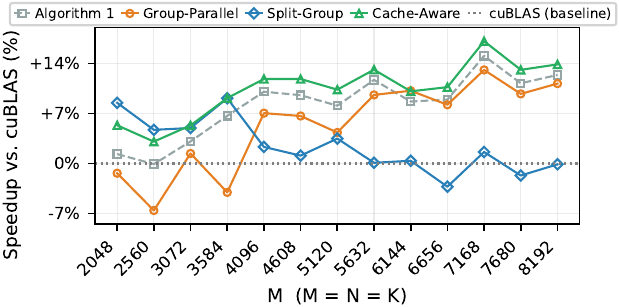}
    \caption{Step-wise evaluation is conducted for the Execution Module on H20 using BF16 precision to run LCMA $(\langle 2,2,2 \rangle, R=7)$. The optimization path is:
             Algorithm~\ref{alg:simplified_workflow} $\to$ Algorithm~\ref{alg:fused_combineA} Group-Parallel $\to$ Split-Group $\to$ Cache-Aware.
            }
    \label{fig:ablation_design}
\end{figure}

To evaluate the effectiveness of the optimizations proposed in Execution Module, we conduct a step-wise evaluation with square matrices on the NVIDIA H20 GPU with BF16 precision. We fix the use of Strassen's algorithm ($\langle 2,2,2 \rangle, R=7$) within \falcongemm{} to simplify the comparison. In Figure~\ref{fig:ablation_design}, \falcongemm{} improves performance over the standard cuBLAS baseline by $3.07\% \sim 17.13\%$ improvement. Our implementation of Algorithm~\ref{alg:simplified_workflow} yields an average speedup of $5.32\%$, and the fusion optimization (Algorithm~\ref{alg:fused_combineA} with Split-Group Parallelism and Cache-Aware Scheduling) achieves an average speedup of $7.83\%$.
However, the Group-Parallel Optimization exhibits unstable performance due to the tail effect caused by its coarse granularity as discussed in Section~\ref{sec:execution}. 
Applying Split-Group Parallelism resolves this and performs better on smaller shapes.
However, its performance drops sharply for $M>4096$, because the increased memory access intensifies L2 cache thrashing and triggers the GPU power limit that drops the SM frequency. 
Finally, the Cache-Aware Scheduling policy mitigates this by reordering computations to increase cache hit rates, delivering stable and high performance that consistently outperforms Algorithm~\ref{alg:simplified_workflow} across all shapes.

\subsection{Roofline Analysis}
\label{subsec:decision_design}

\begin{figure}[t]
    \centering
    \includegraphics[width=0.99\linewidth]{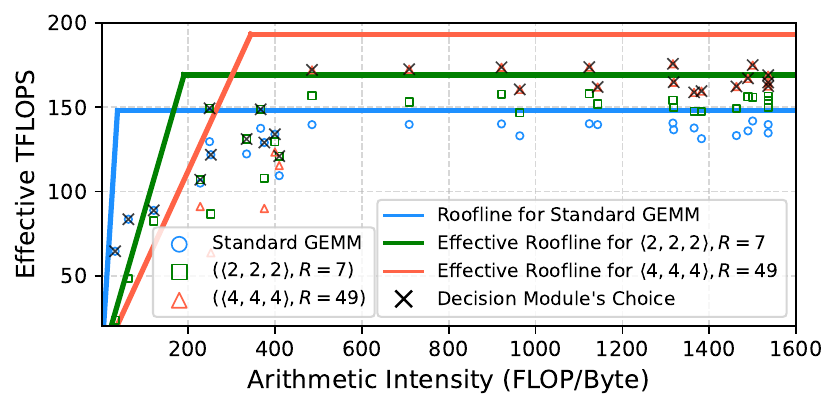}
       \caption{Roofline Analysis on H20 BF16 with 4TB/s bandwidth and 148 TFLOPS.
       This roofline indicates that LCMA pursues higher effective peak TFLOPS at the cost of slightly larger bandwidth. Thus when Arithmetic Intensity is larger enough, LCMA with higher $R$ will be become the optimal solution; otherwise GEMM will be optimal. 
       }
    \label{fig:ablation_model}

\end{figure}

A roofline analysis measured in effective TFLOPS, shown in Figure~\ref{fig:ablation_model}, is used to visualize \falcongemm{}'s LCMA selection strategy proposed by the Decision Module.
We simulate the workloads with varying arithmetic intensities. For high arithmetic intensity workloads, LCMA with a larger $R$ (e.g., $\langle 4,4,4 \rangle, R=49$) yields greater performance gains. But for the extremely low arithmetic intensity ($AI<200$) workloads, LCMA fails to provide benefits as it cannot trade memory bandwidth for computational savings. Standard GEMM becomes optimal solution for these cases.  
Additionally, the peak TFLOPS of LCMA is slightly below the theoretical effective roofline. This minor gap is because extra stages, such as Combine A, cannot be fully overlapped. As a result, the computation units for matrix multiplication cannot operate at peak performance for the entire duration.

\subsection{Numerical Precision Analysis}
\label{subsec:numerical_precision}
A known limitation of LCMA is the reduced numerical stability compared to standard GEMM~\cite{schwartz2024fmmerror,dumas2025towards}. However, our fused pipeline inherently mitigates this issue. Based on our statistical analysis, we observe that \falcongemm{} consistently achieves approximately 17.2\% lower relative error than AlphaTensor across various matrix sizes. Traditional implementations like AlphaTensor typically downcast high-precision intermediate results $H_r$ to lower precision to reduce bandwidth usage when materializing them to off-chip memory. In contrast, \falcongemm{}'s Group-Parallel Optimization fuses the GEMM and Combine H stages. Thus, we calculate the output $C_{i,j}$ directly by using high-precision on-chip $H_r$, yielding better numerical accuracy.

\section{Related Works}
\label{sec:realrelated}
Research on lower-complexity matrix multiplication algorithms (LCMAs) began with Strassen’s rank-7 scheme~\cite{strassen1969gaussian}. Over the following decades, there are numerous theoretical discoveries on new LCMAs~\cite{laderman1990new, alman2025more, williams2024newbounds, dumas2025towards}, such as Laderman's rank-23 algorithm~\cite{laderman1990new}.
While these schemes are typically called \textit{fast matrix multiplication algorithms}, we refer to them as LCMAs to emphasize that lower complexity is not trivially fast and does not always acquire real performance gains on modern hardware.
Recently, AI-driven approaches~\cite{fawzi2022discovering, sun2024opentensor, perminov2026flipgraph} such as AlphaTensor~\cite{fawzi2022discovering}  have discovered new LCMA by novel low-rank decompositions. 

Deploying LCMAs to achieve speedups requires careful hardware-specific implementation.
Benson et al.~\cite{benson2015framework, ballard2012caps} develop an LCMA framework that takes LCMA coefficients as input to generate sequential and distributed-memory parallel implementations.
Dumas et al.~\cite{dumas2025towards} introduce a pipeline that searches for numerically stable LCMA schemes and generates implementations.
AlphaTensor's implementation~\cite{fawzi2022discovering} relies on JAX~\cite{jax2018github} for cross-platform LCMA deployment.
Huang et al. integrate Strassen on CPUs via BLIS~\cite{huang2016strassen} and on NVIDIA GPUs via CUTLASS~\cite{huang2020strassen}, and they also propose a framework to produce LCMA implementations~\cite{huang2017generating}.
Oo and Chaikan~\cite{2023nwe} present a power-efficient Strassen implementation on multi-core CPUs using AVX512 and OpenMP.
Li et al.~\cite{li2011strassengpu} implement Strassen and Winograd matrix multiplication on an NVIDIA GPU.
Krishnan et al.~\cite{gopalakrishnan2021msmes} propose the multi-stage Strassen algorithm for GPUs with no extra GPU memory footprint.
StraGCN~\cite{he2025stragcn} applies Strassen to sparse–dense GNNs using a horizontally fused execution model.
Several of these works~\cite{li2011strassengpu, huang2016strassen, gopalakrishnan2021msmes} also build analytic or empirical performance models to predict when LCMA outperforms classical GEMM.

\section{Conclusion}
\falcongemm{} bridges the long-standing gap between the theoretical potential speedup of LCMAs and their absence in production-level Deep Learning due to a lack of efficient implementation. By addressing the inherent challenges of hardware heterogeneity and deployment complexity, our framework transforms peak-breaking matrix multiplication from a mathematical concept into a viable performance driver for modern Large Language Models.  Our extensive evaluations demonstrate that \falcongemm{} consistently surpasses industry-standard libraries, achieving performance gains of up to 17.85\% and outperforming existing LCMA implementations like AlphaTensor by as much as 55.61\%. These results confirm that \falcongemm{} is a robust, cross-platform solution capable of meeting the rigorous computational demands of LLM training and inference.

\bibliographystyle{IEEEtran}
\bibliography{vince}

\end{document}